\documentclass[conference]{IEEEtran}
\IEEEoverridecommandlockouts
\usepackage{subcaption}
\usepackage[normalem]{ulem}
\usepackage{cite}
\usepackage{amsmath,amssymb,amsfonts}
\usepackage{algorithmic}
\usepackage{graphicx}
\usepackage{textcomp}
\usepackage{xcolor}
\def\BibTeX{{\rm B\kern-.05em{\sc i\kern-.025em b}\kern-.08em
    T\kern-.1667em\lower.7ex\hbox{E}\kern-.125emX}}
\IEEEoverridecommandlockouts
\IEEEaftertitletext{\vspace{-22pt}}
\usepackage[font=small]{caption}
\usepackage{eso-pic}

\begin{document}

\title{\fontsize{22pt}{27pt}\selectfont{Spatial–Frequency Control Authority of
Grid-Forming Inverters over Voltage Waveforms
in Distribution Feeders}}


\author{Amir H. Abolmasoumi and Hamed Mohsenian-Rad,~\IEEEmembership{Fellow,~IEEE}\\
\thanks{The authors are with Winston Chung Global Energy Center, University of California, Riverside, CA, 
USA (e-mails: \{amirhosa, hamedrad\}@ucr.edu).}
}

\maketitle

\AddToShipoutPictureFG*{%
  \AtPageLowerLeft{%
    \put(\LenToUnit{0.35in},\LenToUnit{0.12in}){%
      \parbox{7.8in}{%
        \centering
        \fontsize{5.2}{5.8}\selectfont
        \textcopyright~2027 IEEE. Personal use of this material is permitted.
Permission from IEEE must be obtained for all other uses, in any current
or future media, including reprinting/republishing this material for
advertising or promotional purposes, creating new collective works,
for resale or redistribution to servers or lists, or reuse of any
copyrighted component of this work in other works.
      }%
    }%
  }%
}

\noindent \begin{abstract}
This paper develops a spatial–frequency framework to quantify the impact of distributed energy resources  with grid-forming inverters (GFMs)  on voltage waveforms at different locations along a feeder. An analytical authority function is formulated to characterize how controllability varies with both electrical distance from the point of common coupling (PCC) and frequency. It is shown that the GFM control authority generally decays with both distance and frequency, reflecting the limited ability of inverter actions to influence voltage waveforms at distant points. 
To better capture controllability near resonance, a relative dominance metric is introduced to compare the influence of inverter-induced and grid-induced voltages at a given location and frequency. These results highlight that waveform controllability is fundamentally a spatial–frequency phenomenon along the feeder. Simulation studies validate the analytical findings by demonstrating both the spatial decay of control effectiveness along the feeder and the frequency-dependent trends in control authority in a realistic electromagnetic transient setting. The proposed framework provides insights into the placement and coordination of GFMs to enable effective waveform-level voltage control across electrical distance along the feeder.
\end{abstract}



\vspace{-4mm}
\section{Introduction}

Modern power grids are transitioning toward converter-based systems, introducing new challenges for monitoring and control.
The high penetration of inverter-based resources (IBRs) is altering system dynamics. Fast electromagnetic phenomena, such as sub-cycle oscillations, are becoming more common, especially in weak grids \cite{wideband_oscillation_review, harmonic_resonance_review, PV_oscillation_analysis}. These dynamics span a wide frequency range and are influenced by network impedance and control structure, which cannot be effectively captured using conventional phasor models.

 \vspace{-0.1cm}
\subsection{Motivation: A Practical Challenge}
 \vspace{-0.1cm}

Wide-area control (WAC) enables coordinated regulation of voltage and frequency in large-scale power systems by using synchronized phasor measurements and distributed actuators, such as synchronous generators and, in some cases, IBRs \cite{ PMU_secondary_voltage_control, farahani2023robust}. It is effective for damping inter-area oscillations and supporting secondary voltage control \cite{secondary_voltage_control_survey, wide_area_damping_review_2020}, however, WACs function on electromechanical time scales due to the nature of phasor-measurement-unit (PMU) measurements. PMUs filter out high frequencies such as sub-cycle dynamics. This limits the applicability of WAC to waveform-level disturbances \cite{PMU_applications_review_2025}.

Recent advances in synchronized waveform measurements enable direct observation of instantaneous voltage signals across wide areas using waveform measurement units (WMUs) \cite{synchrowaveforms_2025}. This capability enables the development of control strategies that respond in real time to detailed waveform data, rather than relying on slower, filtered phasor measurements.

\begin{figure}[t]
\vspace{-0.5cm}
    \centering
\includegraphics[width=0.99\linewidth]{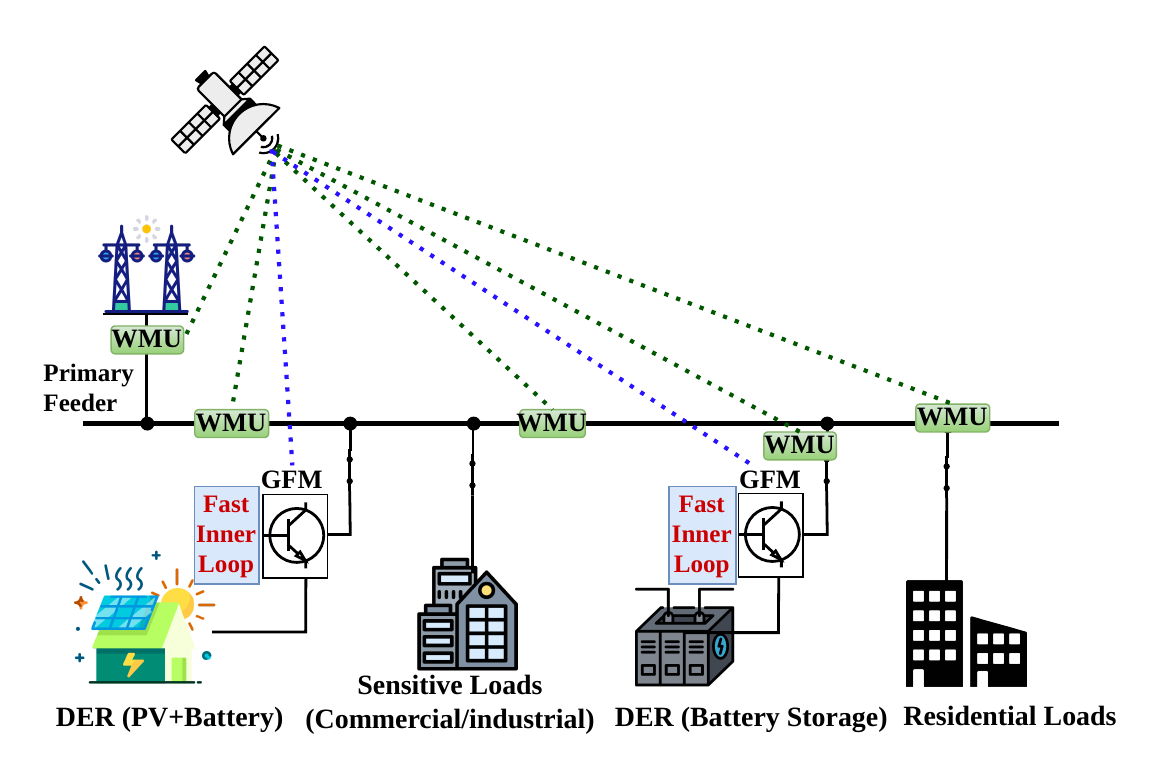}
    \vspace{-0.6cm}
    \caption{Distribution-level waveform-level voltage control using GFMs and  synchronized voltage waveform measurements from WMUs.}
    \vspace{-0.6cm}
    \label{fig:WAEMC}
\end{figure}

Grid-forming inverters (GFMs), which can directly impose voltage waveforms across multiple time scales \cite{khan2024grid, arevalo2025systematic, hakam2026grid}, provide a potential actuation mechanism for wide-area electromagnetic control. Such control leverages the fast inner-loop dynamics of GFMs to adjust voltage waveforms throughout the electrical network. This concept is illustrated in Fig.~\ref{fig:WAEMC}.

However, the success of waveform-level control is restricted by network characteristics. The quality of disturbance propagation through impedances varies with frequency and electrical distance. As a result, the influence of a GFM on the voltage waveforms at desired points could be limited.
This reveals that in IBR-dominated systems, waveform controllability is inherently dependent on spatial location and frequency. Quantifying and analyzing this dependence is a prerequisite for waveform-level control and GFM placement, yet remains largely unexplored. This paper initiates such analysis.
\vspace{-0.1cm}

\subsection{Summary of Contributions}
\vspace{-0.1cm}

 The contributions of the paper can be listed as follows: 
\textbf{(a)} a spatial-frequency analytical framework is developed to quantify the influence of grid-forming inverters on voltage waveforms across a transmission line or distribution feeder. \textbf{(b)} The concept of control authority is introduced to characterize how controllability varies with electrical distance and frequency. \textbf{(c)} The authority ratio is analyzed for two cases of L and LC filters, and the effect of resonance in the LC case is investigated. \textbf{(d)} The empirical authority decay with distance and frequency is verified using detailed EMT simulations.

 \vspace{-0.1cm}

\subsection{Related Literature}
\vspace{-0.1cm}
Recent research on converter-dominated power systems has focused on voltage dynamics beyond the phasor domain, particularly wide-band oscillations and fast electromagnetic interactions. Impedance-based analysis has become a primary approach, where the interaction between converter and grid impedances determines stability over a broad frequency range \cite{wideband_oscillation_review, PV_oscillation_analysis, harmonic_resonance_review}. Various impedance shaping and active damping techniques, such as those in \cite{kuruva2025robust, mukundan2025impedance}, have been proposed to improve stability margins; however, these approaches are primarily oriented toward PCC voltage stabilization rather than explicit controllability analysis of voltage waveforms.


GFMs have been extensively studied for voltage magnitude and angle dynamics \cite{musca2023wide, damping_low_frequency_GFM}, stability, and virtual synchronous machine (VSM) behavior \cite{reviewing_GFM_paradigms, arevalo2025systematic, khan2024grid, hakam2026grid}.

In parallel, advances in synchro-waveform measurements enable direct monitoring of waveform-level dynamics across wide areas \cite{ synchrowaveforms_2025, IBR_waveform_analysis_2025}, while data-driven oscillation estimation techniques provide additional tools for analysis \cite{wideband_oscillation_estimation}.

However, existing works largely focus on stability, local behavior, or monitoring, without explicitly quantifying how voltage control actions propagate spatially. Therefore, the notion of \emph{voltage waveform controllability}, especially its dependence on frequency and electrical distance, remains largely unexplored. This work addresses this gap by developing a spatial–frequency framework to quantify the control authority of GFMs over voltage waveforms along the feeder.

\section{GFM Open-Loop Control Authority \\ Along the Feeder}
In this section, the open-loop influence of the inverter voltage on the network voltage is derived in the Laplace domain. The goal is to obtain a unified expression that quantifies the inverter authority at any electrical distance from the PCC and at any frequency. This formulation provides a spatial and frequency-dependent metric that will later be used to analyze the limitations of electromagnetic control in power systems.
\begin{figure}[t]
    \centering
\includegraphics[width=0.88\linewidth]{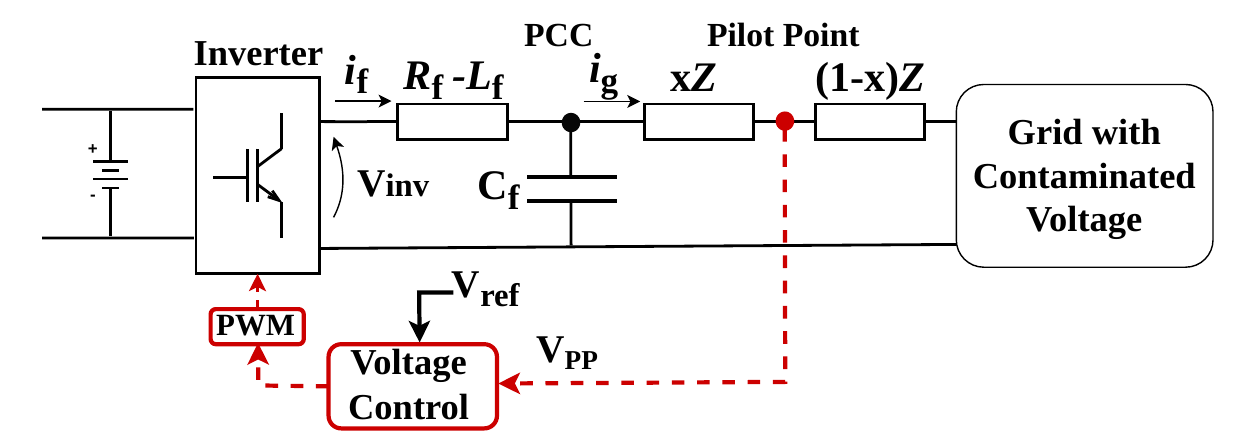}   
\vspace{-0.2cm}
\caption{ Voltage regulation along a feeder by a grid-connected GFM. The target of the voltage waveform control is at the \emph{pilot point} along the feeder, which is electrically away from the GFM's PCC.}
\vspace{-0.6cm}
    \label{fig:Single_GFM}
\end{figure}

We consider the system shown in Fig.~\ref{fig:Single_GFM}. The grid-forming inverter produces a voltage $V_{\mathrm{inv}}(s)$ connected to the PCC through the filter impedance
$Y_f(s)=1/(R_f+sL_f)$. A shunt capacitor admittance $Y_c=sC_f$ is connected at the PCC. The grid voltage $V_g(s)$ is connected through a feeder with the admittance $Y_g=1/(R+sL)$. A pilot point is defined at a fractional electrical distance $x \in [0,1]$ from the PCC.
\vspace{-0.1cm}
\subsection{General Expression of Control Authority Function}


Applying nodal analysis, we can obtain:
 \vspace{-0.1cm}
\begin{equation}
\frac{V_{\mathrm{inv}} - V_{PCC}}{R_f + sL_f}
=
sC_f V_{PCC}
+
\frac{V_{PCC} - V_g}{R + sL}.
\vspace{-0.1cm}
\end{equation}
Solving for $V_{PCC}$ we get
\vspace{-0.2cm}
\begin{equation}
V_{PCC}
=
\frac{{Y_fV_{\mathrm{inv}} + Y_gV_g}}{\Delta(s)},
\vspace{-0.1cm}
\end{equation}
where $\Delta(s)=Y_f(s) + Y_g(s) + Y_c(s)$. This results in
\vspace{-0.1cm}
\begin{equation}
V_{pp}(s,x)=\alpha(s,x)V_{\mathrm{inv}}(s)+\beta(s,x)V_g(s),
\label{eq:1}
\vspace{-0.3cm}
\end{equation}
where
\vspace{-0.2cm}
\begin{align}
\alpha(s,x)= (1-x)\frac{Y_f(s)}{\Delta(s)}, 
\,\, \beta(s,x) = (1-x)\frac{Y_g(s)}{\Delta(s)} + x.
\label{eq:authority}
\vspace{-0.1cm}
\end{align}

\vspace{-0.1cm}
\noindent For a reduced representation of local GFM control,
\vspace{-0.2cm}
\begin{equation}
V_{\mathrm{inv}}
=
K_v(s)\bigl(V^\star-V_{PCC}\bigr)-K_i(s)i_f .
\label{eq:local_gfm}
\vspace{-0.2cm}
\end{equation}
The resulting closed-loop pilot-point voltage can be expressed as $V_{pp}=\alpha_{\mathrm{eff}}V^\star+\beta_{\mathrm{eff}}V_g$
where
\vspace{-0.2cm}
\begin{equation}
  \alpha_{\mathrm{eff}}
=
\alpha\frac{K_v}{D_c}, \,\, \beta_{\mathrm{eff}}
=
\beta+
\alpha\frac{Y_g}{\Delta}
\frac{-K_v+K_iY_f}{D_c},
\label{eq:effective_authority}
\vspace{-0.1cm}  
\end{equation}
\noindent with $D_c=1+K_vH+K_iY_f(1-H),$ and $H=Y_f/\Delta$. Local voltage and current loops reshape the effective contributions, while $\alpha$, $\beta$ capture network coupling; virtual impedance further modifies it through $Y_f$.

\subsection{Spatial–Frequency Dependence of GFM Authority}
\vspace{-0.1cm}

\begin{figure}[t]
    \centering 
    \begin{subfigure}{0.24\textwidth}
        \includegraphics[width=\textwidth]{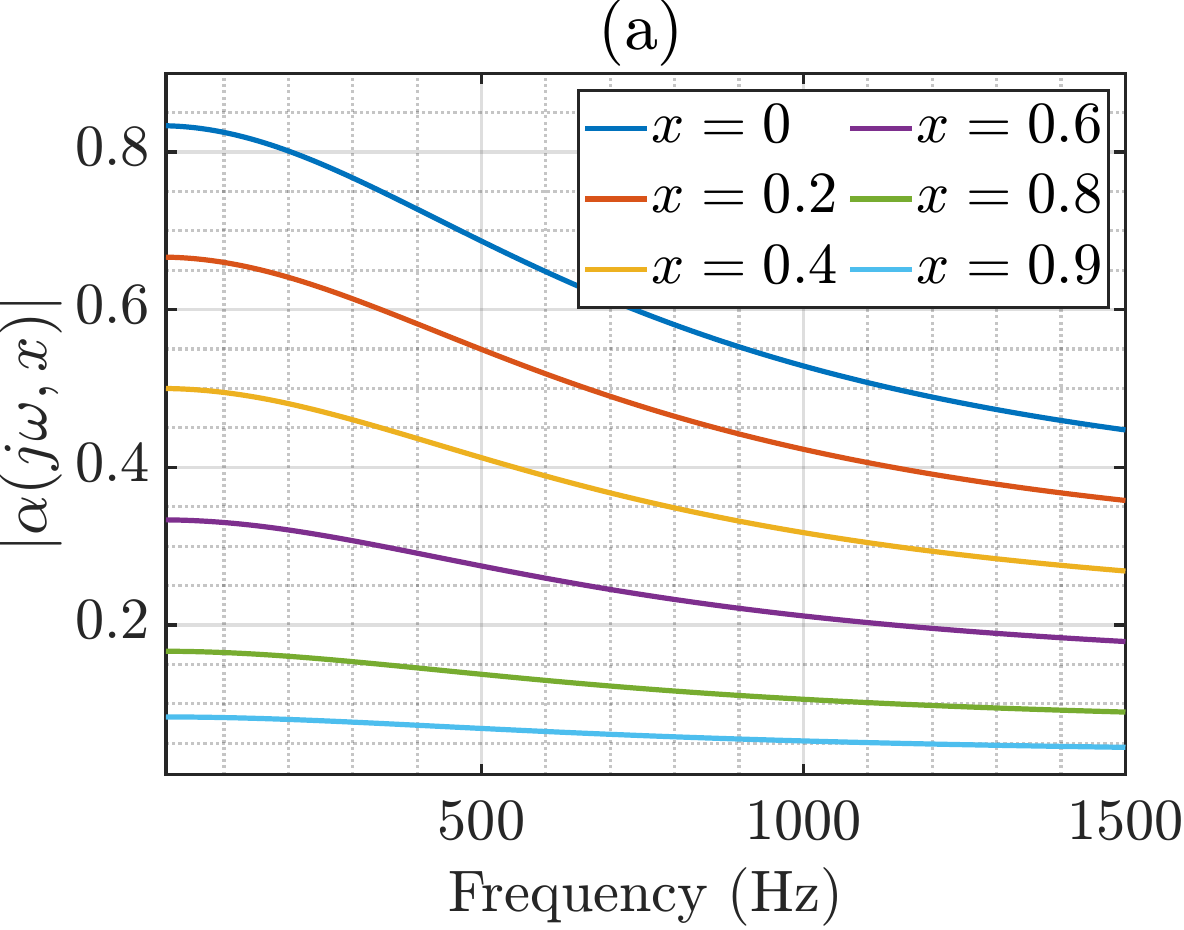} 
        \label{fig:Alpha_low}
    \end{subfigure}
    \begin{subfigure}{0.24\textwidth}
        \includegraphics[width=\textwidth]{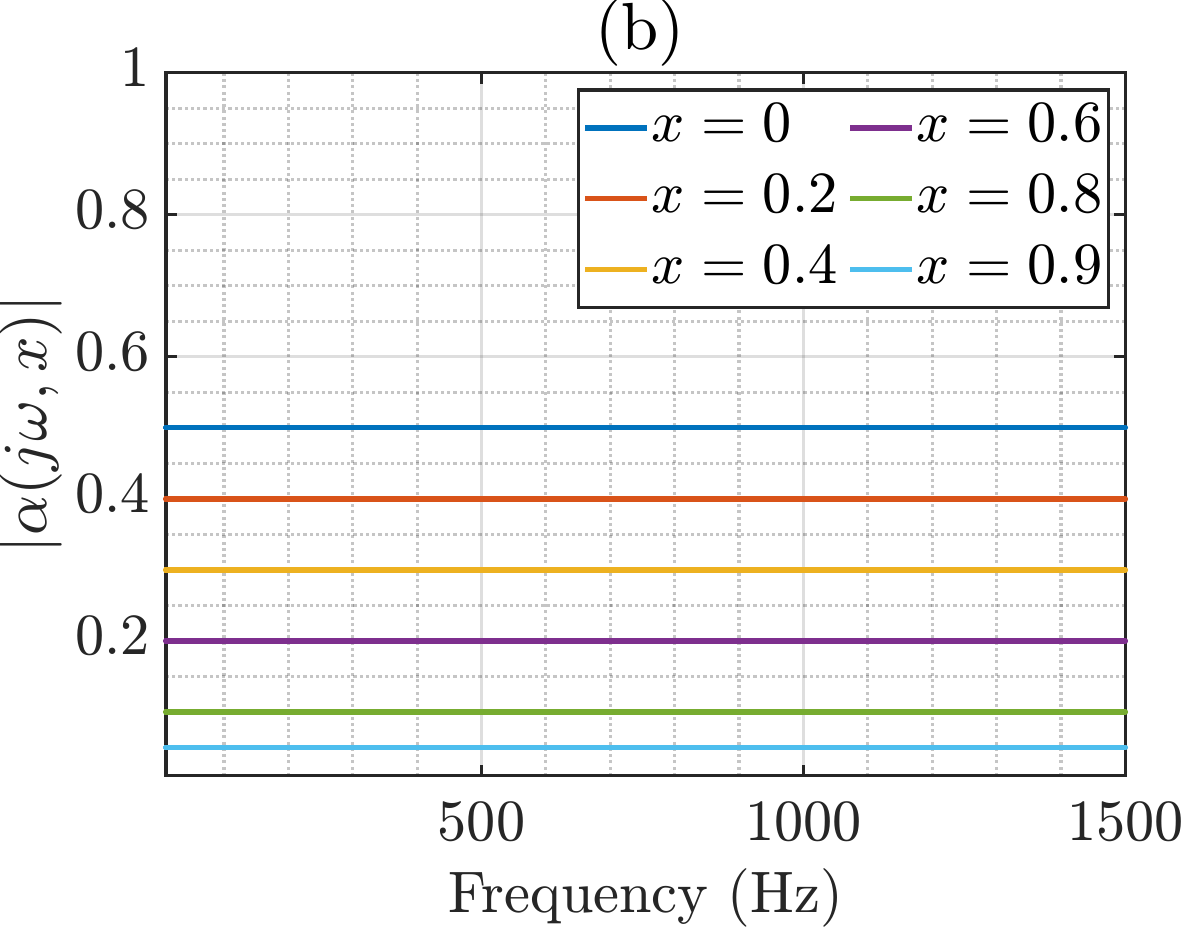} 
        \label{fig:Alpha_large}
        \end{subfigure}          \vspace{-0.7cm}  \\
    \begin{subfigure}{0.24\textwidth}
        \includegraphics[width=\textwidth]{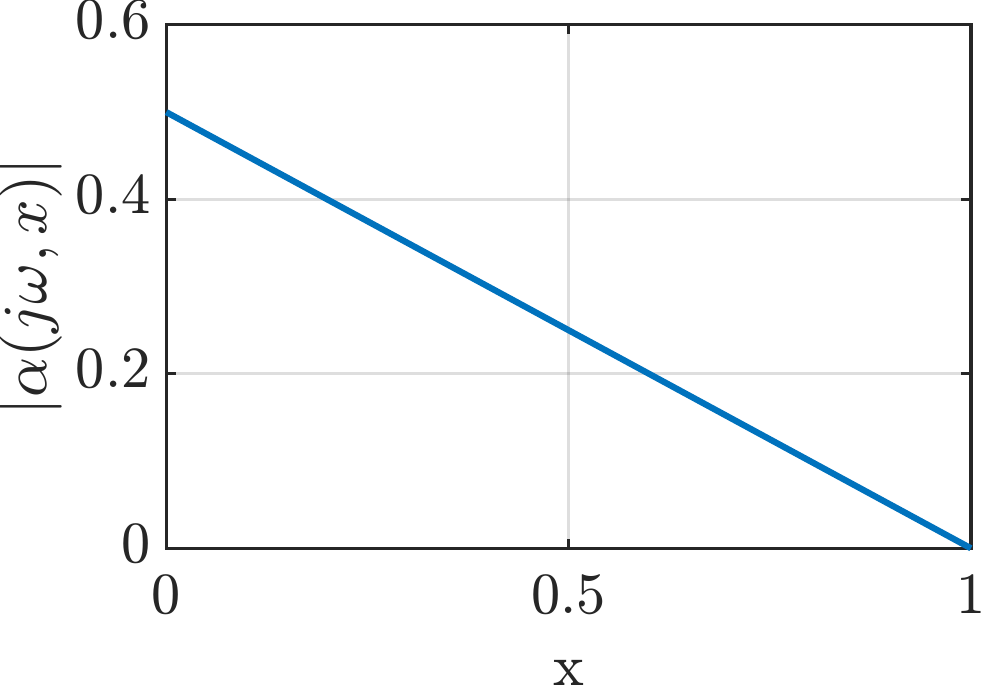} 
        \label{fig:Alpha_x}
    \end{subfigure}
    \vspace{-0.5cm}
    \caption{Spatial-frequency behavior of Authority function; (a) Small L/R, (b) Large L/R, and (c) Decay with distance at 500Hz.}
    \label{fig:Alpha_analysis}
     \vspace{-0.5cm}
\end{figure}

The magnitude of the control authority function in absence of the shunt capacitance can be obtained as:
\vspace{-0.1cm}
\begin{equation}
|\alpha(j\omega,x)|^2 =
(1-x)^2
\frac{R^2+\omega^2L^2}{(R+R_f)^2+\omega^2(L+L_f)^2},
\label{eq:alpha_mag}
\vspace{-0.1cm}
\end{equation}
where $\omega$ denotes the angular frequency and $x\in[0,1]$ represents the electrical distance from the PCC. In this case $\beta=1-\alpha$. Equation \eqref{eq:alpha_mag} shows that inverter authority is determined by the spatial factor $(1-x)$, representing attenuation along the feeder, and the electromagnetic interaction between the feeder and inverter filter impedances. 
The frequency dependence of \eqref{eq:alpha_mag} arises from the relative magnitude of the resistive and inductive terms.
If the feeder is sufficiently inductive such that for middle frequencies, 
say $f \in [200,1000]~\mathrm{Hz}$,
we have $\omega L \gg R$, then neglecting the resistive terms in \eqref{eq:alpha_mag} leads to
\vspace{-0.2cm}
\begin{equation}
|\alpha(j\omega,x)| \approx
(1-x)\frac{L}{L+L_f}.
\label{eq:inductive_result}
\vspace{-0.2cm}
\end{equation}
Therefore, for inductive feeders, inverter authority depends solely on the spatial distance and the inductive impedance division between the feeder and inverter filter. 
For less inductive feeders, particularly if ${L}/{R}<{L_f}/{R_f}$, the authority would decrease with both frequency and distance within the middle frequency band.
These results are shown in Fig \ref{fig:Alpha_analysis}. As seen, for highly inductive lines the decay in authority is mostly spatial but for less inductive lines the authority decreases with both distance and frequency. From control viewpoint, this implies that control of high-frequency oscillations on distant points needs less control effort for inductive systems  while it may need higher control effort for the resistive grids as both spatial and frequency decays should be addressed.
\subsection{Intermediate-Frequency Behavior with Shunt Capacitance}
The previous section established the spatial and frequency dependence of inverter authority without shunt capacitance, showing that it either decreases monotonically with frequency or becomes nearly frequency independent for sufficiently inductive feeders. In practice, however, shunt capacitance from inverter filters, cables, or reactive compensation modifies the inverter-grid interaction and introduces resonant dynamics. With the shunt capacitor $C_f$ included at the PCC, \eqref{eq:1} remains valid, but $\alpha+\beta\neq1$.

By putting admittance terms in \eqref{eq:authority} we obtain:
\vspace{-0.2cm}
\begin{equation}
\begin{aligned}
    \alpha(s,x) &= \frac{(1-x)(R+sL)}{D(s)}, \\
    D(s)   &= LL_fC_f s^3 + (R_fLC_f + RL_fC_f)s^2 + \\
                & \quad (L + L_f + RR_fC_f)s + (R + R_f).
\end{aligned}
\label{eq:alpha_polynomial}
\vspace{-0.1cm}
\end{equation}
From \eqref{eq:alpha_polynomial}, the shunt capacitor introduces LCL-type dynamics. The approximate resonance frequency is given by
\begin{equation}
\omega_r^2\approx\frac{L+L_f+RR_fC_f}{LL_fC_f}.
\label{eq:res_frequency}
\vspace{-0.1cm}
\end{equation}

For a typical range of GFM filter and feeder parameters, namely $L_f\approx1$--$3,\mathrm{mH}$, $L\approx1$--$5,\mathrm{mH}$, $R_f\approx0.02$--$0.1,\Omega$, $R\approx0.2$--$1,\Omega$, and $C_f\approx10$--$40,\mu\mathrm{F}$, \eqref{eq:res_frequency} predicts approximate resonance frequencies of $f_r\approx0.8$--$1.8,\mathrm{kHz}$. This range is close to the upper end of the waveform-control band, $200Hz-1kHz$, indicating that resonant dynamics may influence its operation.
As an example, consider the parameter set $L_f=2\,\mathrm{mH},
L=2\,\mathrm{mH},$ and $C_f=30\,\mu\mathrm{F}.$
Substituting these values into \eqref{eq:res_frequency} gives $f_r=\omega_r/(2\pi)\approx920,\mathrm{Hz}$, which lies near the upper end of the waveform-control band.

As a result, resonant dynamics of practical GFM-grid interconnections can influence the frequency range targeted for waveform control. Near $\omega_r$, reactive cancellation reduces the total admittance at the PCC, causing $|\alpha(j\omega,x)|$ to exhibit a resonance peak. Consequently, the monotonic decay or nearly frequency-independent behavior observed for $C_f=0$ is replaced by an increase in the pre-resonant region. Despite this local increase, the spatial attenuation factor $(1-x)$ remains unchanged. Thus, resonance locally modifies the frequency dependence while preserving the distance dependence.


From~\eqref{eq:1}, voltage regulation at the pilot point can be interpreted as a competition between the inverter voltage $V_{\mathrm{inv}}$ and the grid voltage $V_g$, which includes the disturbance. The magnitude of $\alpha$ measures the sensitivity of $V_{pp}$ to $V_{\mathrm{inv}}$, but not its dominance. Near resonance, both $\alpha$ and $\beta$ may increase due to the small denominator in \eqref{eq:alpha_polynomial}. Therefore, controllability is better characterized by the relative dominance of the inverter and grid contributions, defined as
\vspace{-0.2cm}
\begin{equation}
\Gamma(s,x)=\frac{\alpha(s,x)}{\beta(s,x)}.
\label{eq:gamma_def}
\vspace{-0.1cm}
\end{equation}
Using \eqref{eq:1}, this ratio simplifies to
\vspace{-0.1cm}
\begin{equation}
\Gamma(s,x)
=
\frac{(1-x)Y_f(s)}
{xY_f(s)+Y_g(s)+xY_c(s)}.
\label{eq:gamma_simplified}
\vspace{-0.1cm}
\end{equation}
\begin{figure}[t]
    \centering
\includegraphics[width=0.56\linewidth]{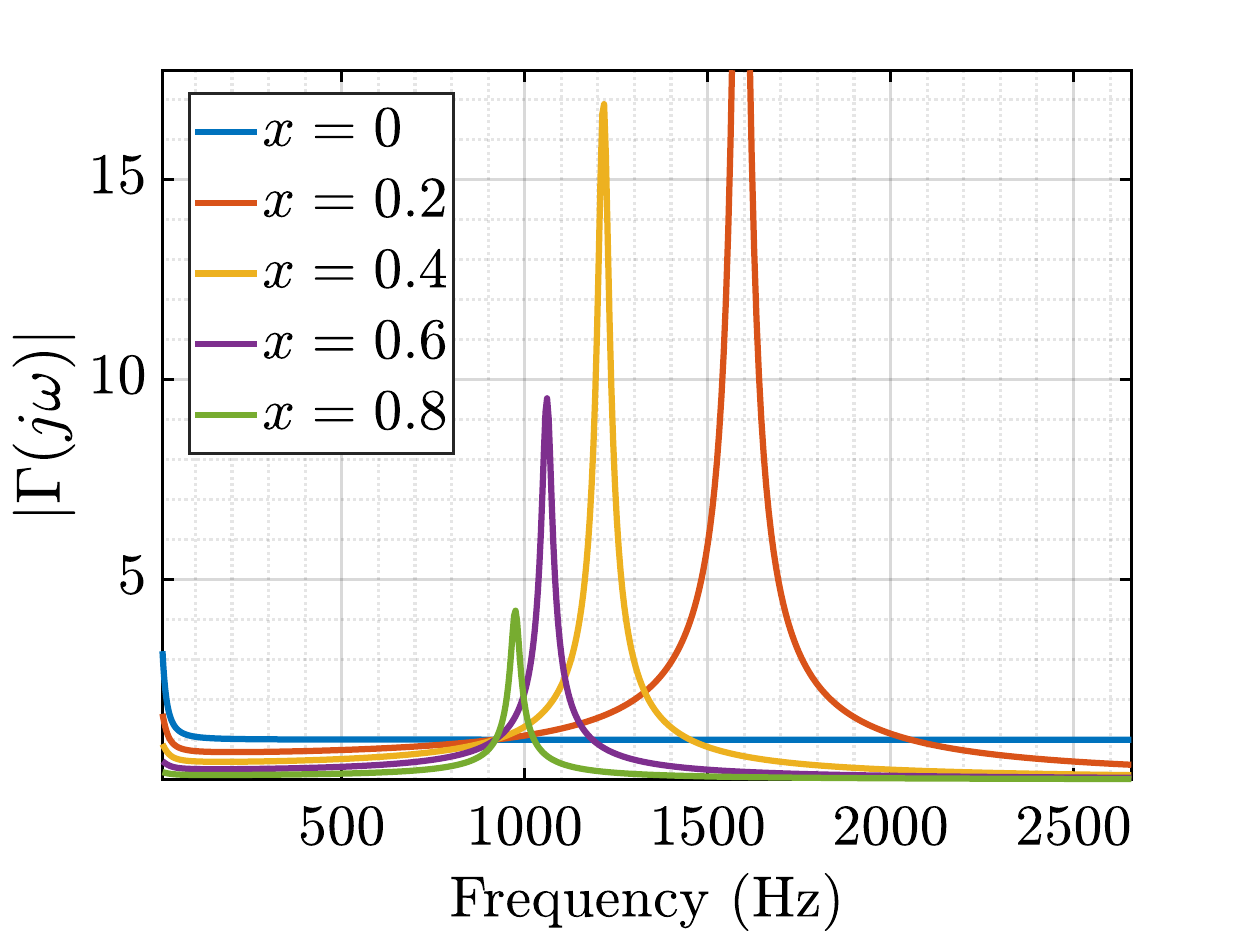} 
\vspace{-0.1cm}
\caption{Relative control authority dominance ratio.}
    \label{fig:relative_authority}
\vspace{-0.6cm}
\end{figure}
Unlike $\alpha$, the metric $\Gamma$ directly compares inverter-induced voltage to grid-induced voltage at the same frequency. When $|\Gamma(j\omega,x)| > 1$, the inverter contribution dominates; when $|\Gamma(j\omega,x)| < 1$, the grid dominates.
Figure~\ref{fig:relative_authority} shows $|\Gamma(j\omega,x)|$, which exhibits a pronounced peak near the resonant region. This peak results from resonance in the reduced admittance $xY_f+Y_g+xY_c$, whose imaginary part becomes zero at
\vspace{-0.1cm}
\begin{equation}
\omega_{r,\Gamma}^2 \approx \frac{L_f + x L}{x L L_f C_f}.
\vspace{-0.1cm}
\end{equation}
This resonance differs from that of the authority function $|\alpha|$, as it arises from the reduced admittance $xY_f+Y_g+xY_c$ rather than the total admittance. Physically, it identifies a narrow frequency band where the relative dominance of the inverter contribution increases locally. 
For $x>0$, the resonance frequency increases as $x$ decreases. At $x=0$, the capacitor term cancellation leads to no resonance.
Moreover, the peak becomes larger and shifts in frequency. 
From \eqref{eq:gamma_simplified}, as $x$ decreases, the denominator approaches $Y_g$, increasing the relative dominance of $Y_f$.

For a target frequency set $\Omega$, candidate GFM locations can be compared using $\mathcal{A}_{\Gamma}(x)
=\min_{\omega\in\Omega}|\Gamma(j\omega,x)|$.
Larger $\mathcal{A}_{\Gamma}$ indicates stronger authority over the target frequency range. For multiple GFMs and pilot points, the concept extends to combined authority for placement and coordination.

\begin{figure}[t]
    \centering 
\includegraphics[width=0.43\textwidth]{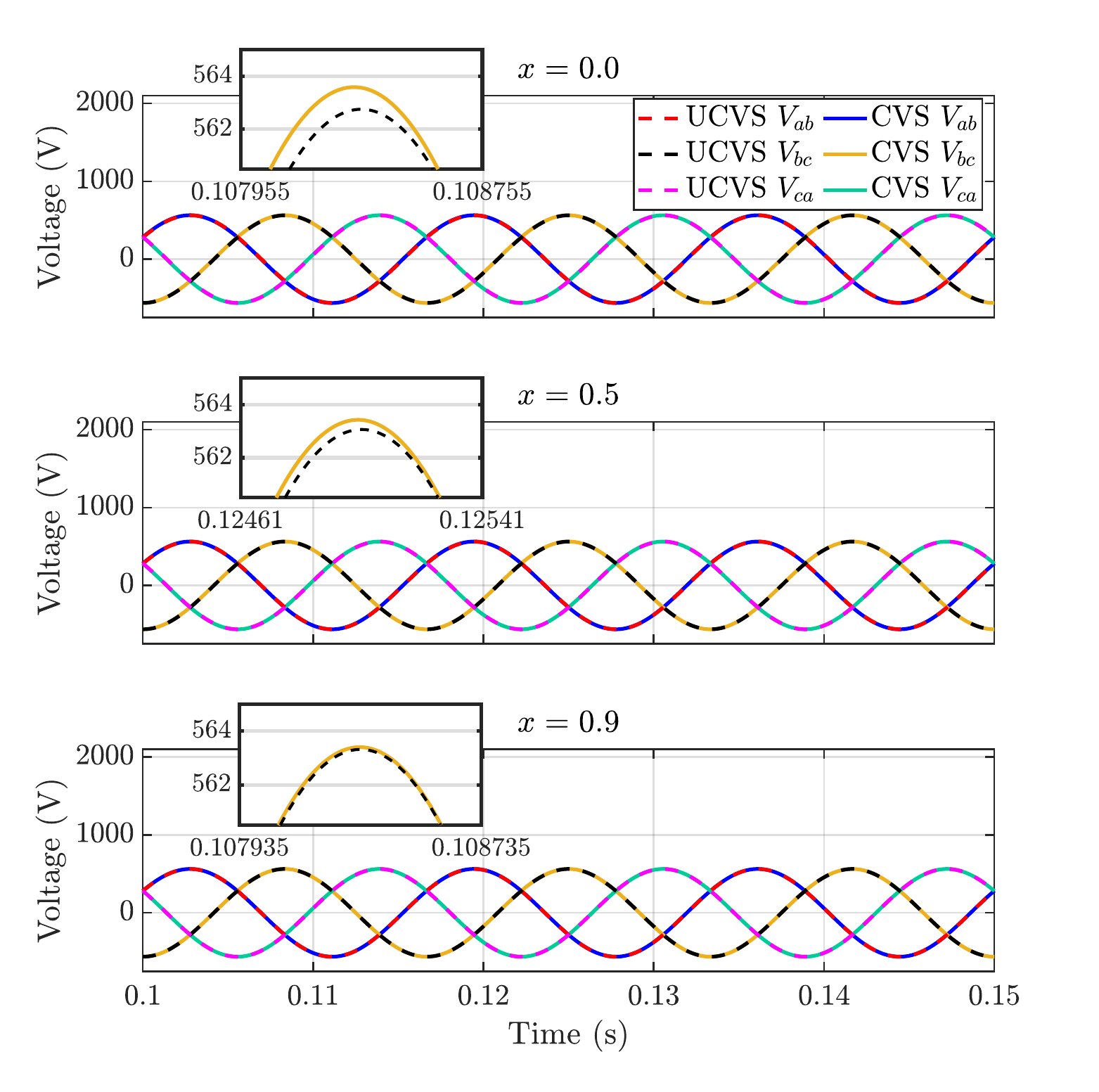} 
    \vspace{-0.3cm}\caption{Voltage tracking across the
feeder for a disturbance-free scenario. The feedback signal is received from the PCC.}
    \vspace{-0.3cm}\label{fig:Simu1_Voltages}
\end{figure}
\begin{figure}[t]
    \centering 
    \includegraphics[width=0.32\textwidth]{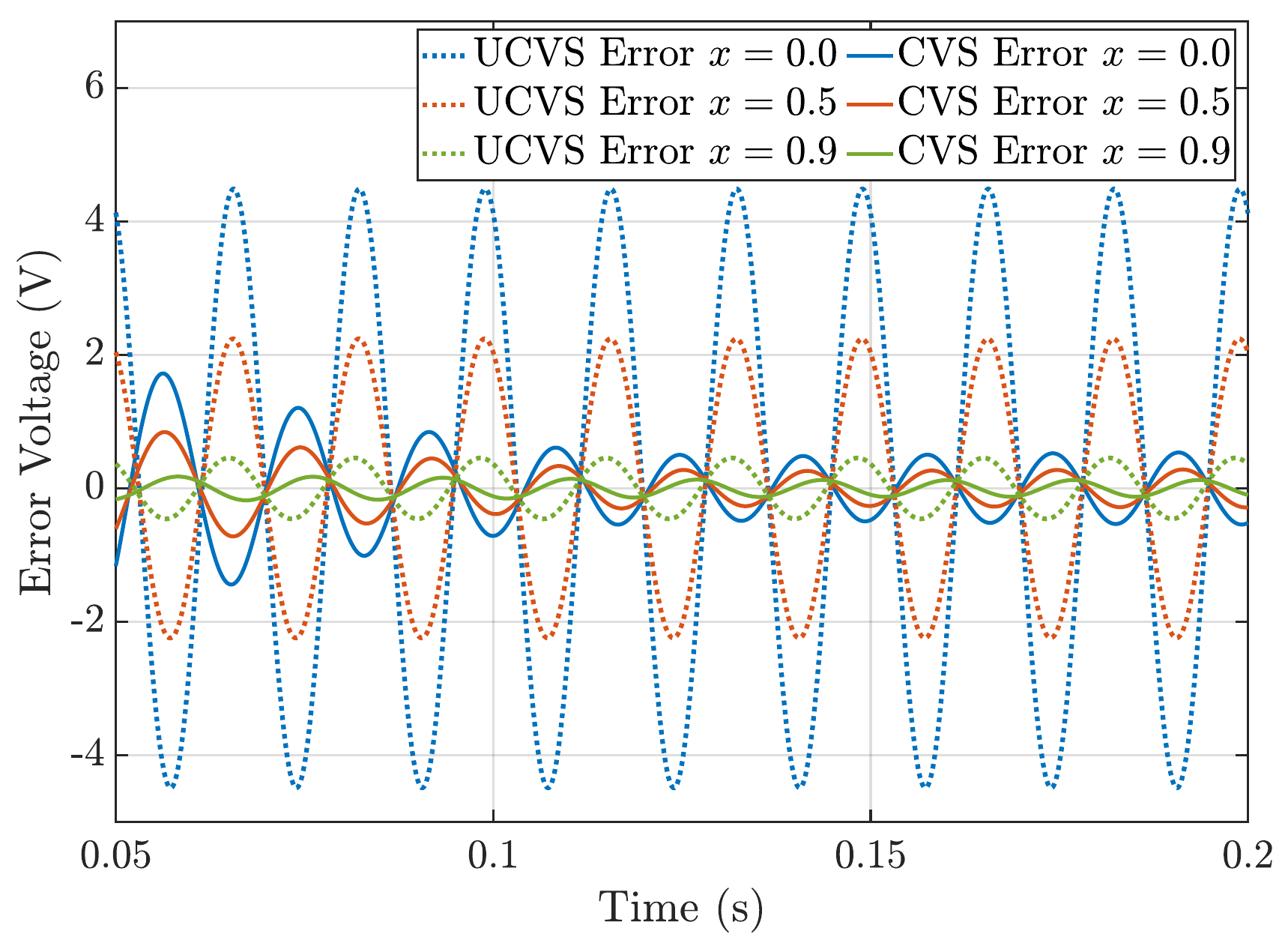} 
    \vspace{-0.1cm}
    \caption{Voltage regulation errors for different choices of the electrical distance ratio x. Only phase A is shown here.}
    \vspace{-0.4cm}\label{fig:Simu1_Voltage_Errors}
\end{figure}

\begin{figure}[h]
    \centering 
    \includegraphics[width=0.32\textwidth]{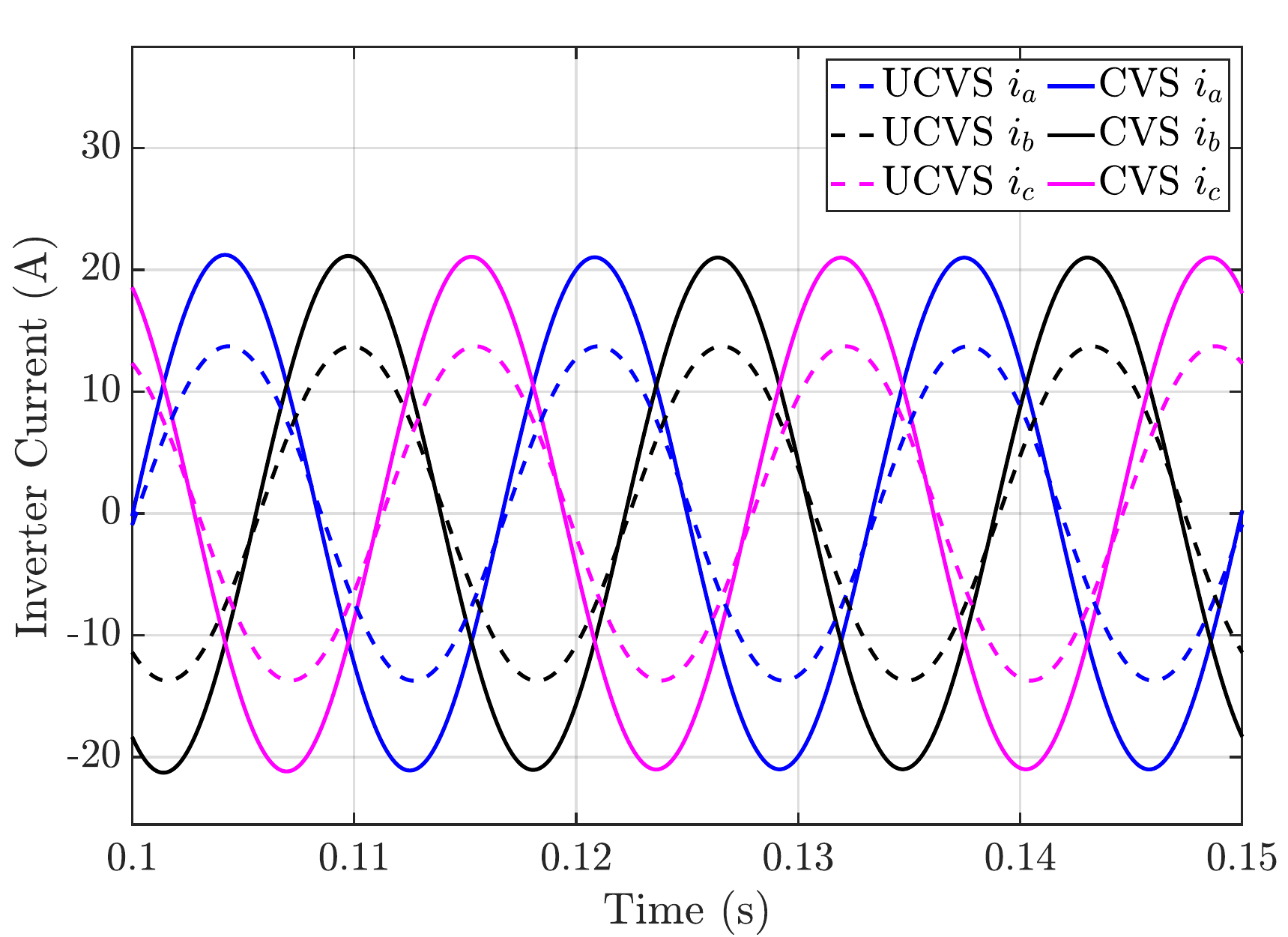} 
    \vspace{-0.2cm}
    \caption{Currents injected by the uncontrolled and controlled GFM.}
    \label{fig:Simu1_Currents}
    \vspace{-0.1cm}
\end{figure}
\begin{figure}[h]
    \centering 
    \includegraphics[width=0.32\textwidth]{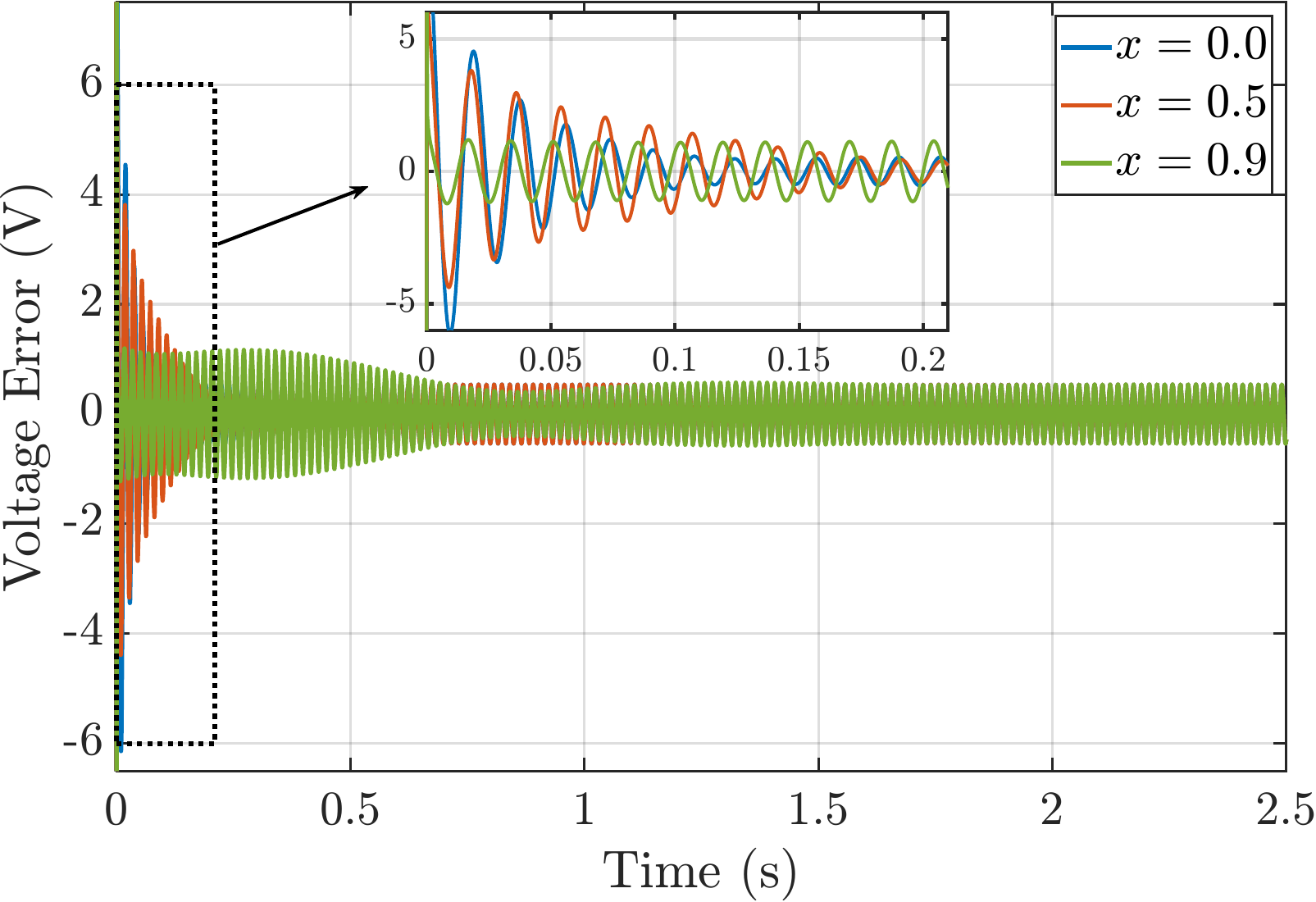} 
    \vspace{-0.1cm}
    \caption{Voltage tracking errors taking feedback from pilot point at x.}
    \label{fig:Voltage_Tracking_Error}
    \vspace{-0.5cm}
\end{figure}
\vspace{-0.2cm}
\begin{figure}[h]
    \centering 
    \includegraphics[width=0.32\textwidth]{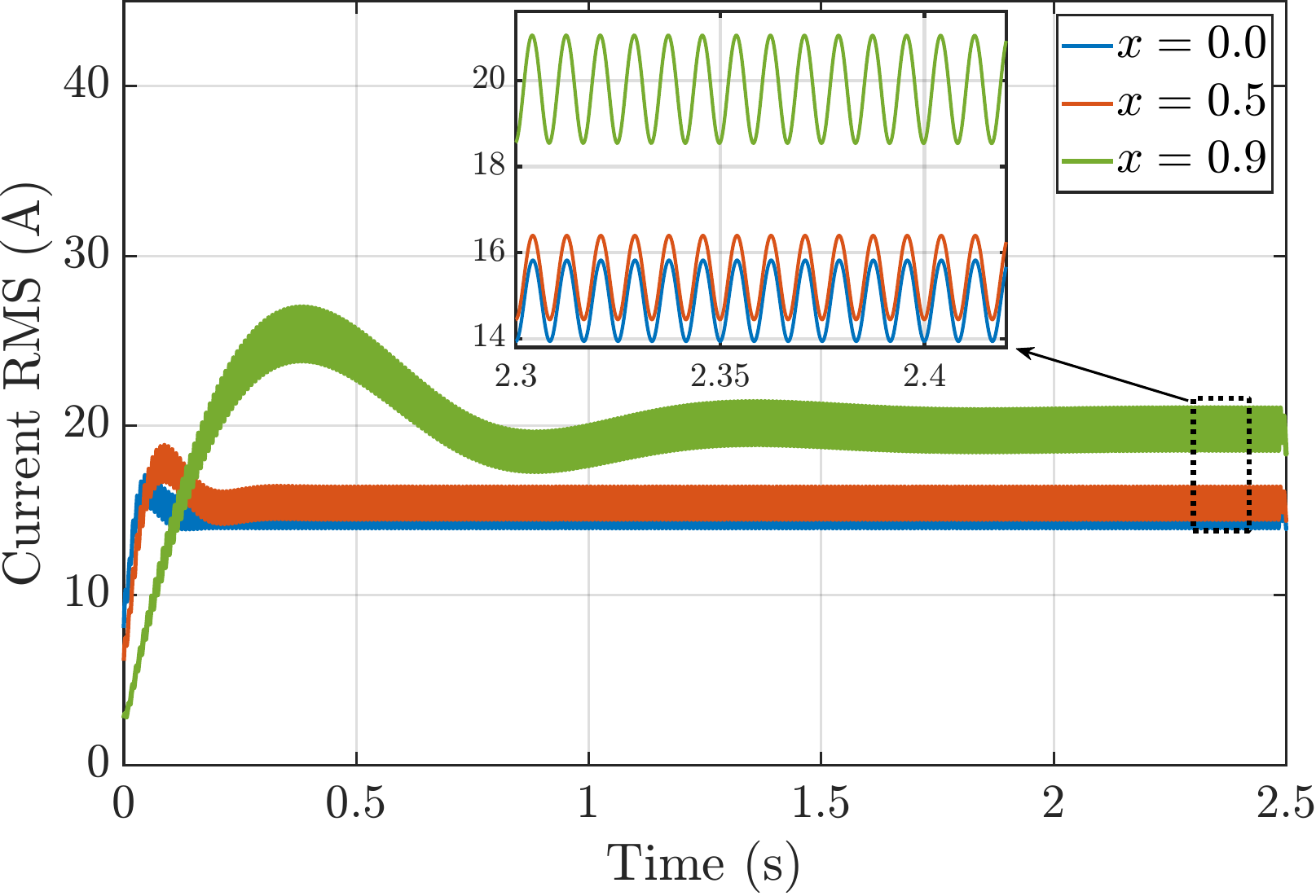} 
   \vspace{-0.1cm}
    \caption{currents injected by a GFM that takes the feedback from pilot point at x along the feeder using a  WMU, instead of from the PCC. }
    \label{fig:Inverter_Current_FeedbackfromX}
    \vspace{-0.4cm}
\end{figure}
 \vspace{0.2cm}
\section{Simulation Results}

The EMT simulations are performed in MATLAB/Simulink Simscape Electrical Specialized Power Systems (SPS)~\cite{matlab_sps}, where the circuit diagram of Fig.~\ref{fig:Single_GFM} is constructed and tested. 
\vspace{-0.1cm}
\subsection{Spatial Decay of GFM Control in Disturbance-free Case}
This section investigates the voltage regulation at the PCC and along the feeder to evaluate the spatial reach of a GFM unit. A cascaded PI controller regulates the PCC voltage using feedback from PCC voltage and inverter current. In such controller the voltage loop provides the reference for faster current loop and the PI control gains are designed appropriately to have reasonable transient response at PCC. Two cases are compared: an uncontrolled voltage source (UCVS) and a controlled voltage source (CVS) with mentioned vector PI controller. A resistive load $R=16\,\Omega$ is connected at the PCC.
Results in Fig.~\ref{fig:Simu1_Voltages}, \ref{fig:Simu1_Voltage_Errors} show that at $x=0$ (PCC), the CVS achieves smaller tracking error than the UCVS. As we move towards the grid ($x \to 1$), the error decreases and becomes almost equal for both CVS and UCVS near the grid where the voltage is dominated by the stiff grid. This demonstrates the spatial decay of GFM control authority regardless of how $V_{inv}$ is generated (UCVS or CVS). Fig.~\ref{fig:Simu1_Currents} shows that the CVS requires higher current injection than the UCVS, reflecting increased control effort to regulate PCC voltage under load and feeder impedance. In practice, GFM inverters are current-limited and large loads may push operation toward thermal or hardware limits, requiring careful control design.


Next, voltage feedback is taken from a location $x \in [0,1]$ along the feeder (Fig.~\ref{fig:Single_GFM}). In practice, this requires a transmitter sensor and synchronized measurements using WMUs \cite{synchrowaveforms_2023}. The voltage tracking errors in Fig.~\ref{fig:Voltage_Tracking_Error} show that the transient response depends strongly on the feedback location. At the PCC ($x = 0$), the response is well damped. The larger initial error reflects a local transient mismatch rather than low controllability. For $x = 0.5$, both the magnitude and decay rate become moderate due to the combined influence of the GFM and the grid. At $x = 0.9$, the initial error is smaller due to grid dominance; however, the response decays more slowly, indicating the inverter’s limited ability to shape remote voltage waveforms. All cases converge to small steady-state errors.

Fig.~\ref{fig:Inverter_Current_FeedbackfromX} shows that the inverter current increases with $x$. This directly indicates a higher effort to regulate distant points. 
However, such high effort does not improve performance; for example, for $x=0.9$, a larger current corresponds to slower error decay. 
This behavior is consistent with the authority function $\alpha(s,x)$ (for $C_f=0$), which decreases with distance, reducing inverter influence relative to the grid.
\vspace{-0.1cm}
\subsection{Empirical Authority Decay with Frequency}

To validate the proposed theoretical framework, the Control Authority Identification method is implemented in an EMT simulation environment. This empirical approach characterizes the physical coupling using the following decomposition:
\vspace{-0.1cm}
\begin{equation}
{{\alpha}_{Emp}}(j\omega, x) = \frac{\mathbf{V}_{pp}(\omega) - \mathbf{V}_{g}(\omega)}{\mathbf{V}_{inv}(\omega) - \mathbf{V}_{g}(\omega)}
\vspace{-0.1cm}
\end{equation}

\begin{figure}[t]
    \centering 
    \includegraphics[width=0.32\textwidth]{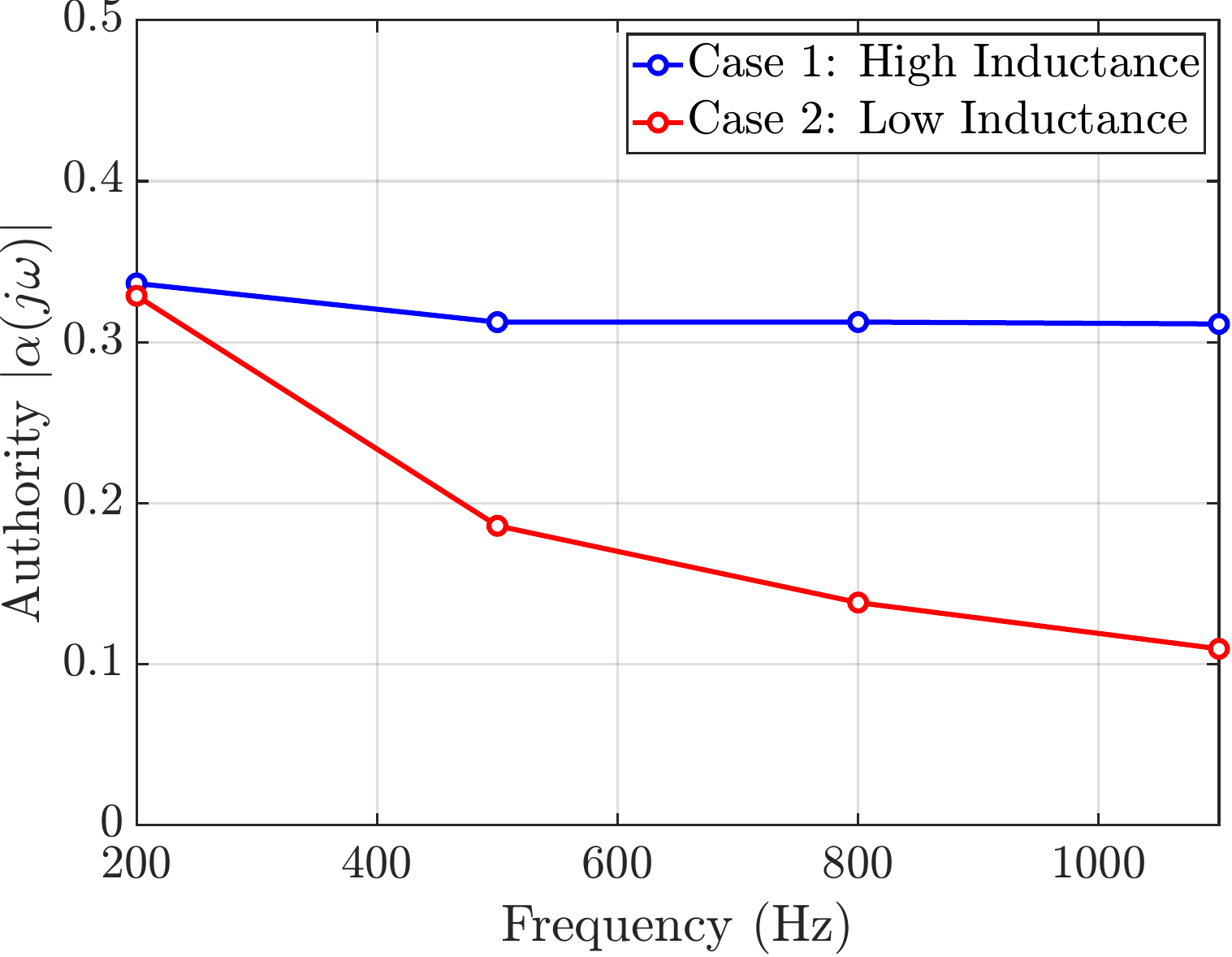} 
    \vspace{-0.1cm}
    \caption{The empirical alpha decays with increase in frequency. The decay also depends on the impedance, which itself is a function of electrical distance between the PCC and pilot point. This creates a spatial-frequency dependency that characterizes the control authority.
    }
    \label{fig:Empirical_alpha}
    \vspace{-0.6cm}
\end{figure}

The complex phasors $\mathbf{V}(\omega)$ are extracted from time-domain EMT waveforms using a synchronized Fourier projection. As shown in Fig.~\ref{fig:Empirical_alpha}, for a constant $x = 0.5$, the magnitude $|\alpha|$ generally decays as frequency increases. This decay is more pronounced for low-inductance lines and remains nearly constant for highly inductive lines. These results match the theoretical analysis of control authority, confirming that coupling reactance dominates the control reach in inductive networks.

\section{Conclusions and Future Work}

This paper developed a spatial–frequency framework to quantify the control authority of GFMs over voltage waveforms at pilot points along a feeder. A closed-form authority function was derived to characterize how inverter influence varies with both electrical distance and frequency. The analysis showed that control authority generally decays with distance and, for low $L/R$ feeders, also with frequency. 
Shunt capacitance was shown to introduce resonant dynamics that locally amplify authority. A relative dominance metric was introduced to better characterize controllability near resonance. EMT simulations validated the framework by confirming spatial decay of control effectiveness, increased control effort at distant points, and frequency-dependent authority trends. These results demonstrate that waveform-level voltage controllability is inherently a spatial–frequency phenomenon and establish a practical foundation for analyzing the reach and limitations of GFMs in waveform-level control.

%

Future work will extend the analysis to multi-GFM networks with multiple pilot points and interacting feeders, enabling the characterization of controllability across frequency and electrical distance. Several challenges remain. These include scalable modeling of spatial–frequency controllability, coordination among multiple converters, and communication and synchronization for remote feedback using WMUs. 
Addressing these challenges is essential to enable practical electromagnetic control of voltage in IBR-dominated systems.
\bibliographystyle{IEEEtran}
\vspace{-0.1cm}
\bibliography{Ref}

@article{PMU_secondary_voltage_control,
  author={Pierrou, Georgia and Wang, Xiaozhe},
  journal={IEEE Transactions on Power Systems}, 
  title={An Online Network Model-Free Wide-Area Voltage Control Method Using PMUs}, 
  year={2021},
  volume={36},
  number={5},
  pages={4672-4682}
  }

@article{farahani2023robust,
  author={Ali Farahani and Amir H. Abolmasoumi and Lamine Mili and Mohammad Bayat},
  title={A Robust Fusion Bus Frequency Estimation Method to Improve Frequency Oscillation Damping in Power Systems},
  journal={IET Generation, Transmission \& Distribution},
  volume={17},
  number={18},
  pages={4172--4187},
  year={2023},
  doi={10.1049/iet-gtd.2023.0188}
}

@article{secondary_voltage_control_survey,
  author={H. Bevrani and T. Hiyama and Y. Mitani},
  title={Survey of the Secondary Voltage Control: Realization and Investigations},
  journal={Electric Power Systems Research},
  volume={92},
  pages={30--38},
  year={2012},
  doi={10.1016/j.epsr.2012.02.015}
}

@article{wide_area_damping_review_2020,
  author={H. Singh and N. Kishor and A. K. Singh},
  title={Wide-Area Damping Control in Renewable Integrated Power Systems: A Review on Recent Achievements and New Challenges},
  journal={Engineering},
  volume={6},
  pages={695--714},
  year={2020},
  doi={10.1049/enc2.70027}
}

@article{PMU_applications_review_2025,
  author={Z. Li and Y. Wang and L. Chen},
  title={Comprehensive Review of {PMU} Applications in Smart Grid: Enhancing Grid Reliability and Efficiency},
  journal={Chinese Journal of Electrical Engineering},
  volume={11},
  number={1},
  pages={129--150},
  year={2025},
  doi={10.23919/CJEE.2025.000129}
}

@article{wideband_oscillation_review,
  author={Runzhi Mu and Yuming Zhang and Xiongbiao Wan and Deng Wang and Tianshu Wen and Liming Sun and Zichao Zhou and Bo Yang},
  title={Analysis and Mitigation of Wideband Oscillations in {PV}-Dominated Weak Grids: A Comprehensive Review},
  journal={Processes},
  volume={13},
  number={11},
  pages={3450},
  year={2025},
  doi={10.3390/pr13113450}
}

@article{harmonic_resonance_review,
  author={Jinggeng Gao and Yong Yang and Honglei Xu and Yingzhou Xie and Chen Zhou and Haiying Dong},
  title={The Wideband Oscillatory Localization Method Based on Combining Compressed Sensing and Graph Attention Networks},
  journal={Energies},
  volume={17},
  number={23},
  pages={6062},
  year={2024},
  doi={10.3390/en17236062}
}

@article{PV_oscillation_analysis,
  author={L. Fan and Z. Miao and D. Piper and P. Ramasubramanian and L. Zhu and P. Mitra},
  title={Analysis of 0.1-{Hz Var} Oscillations in Solar Photovoltaic Power Plants},
  journal={IEEE Transactions on Sustainable Energy},
  volume={14},
  pages={734--737},
  year={2023},
  doi={10.1109/TSTE.2022.3206795}
}

@article{synchrowaveforms_2023,
  author={Hamed Mohsenian-Rad and Wilsun Xu},
  title={Synchro-Waveforms: A Window to the Future of Power Systems Data Analytics},
  journal={IEEE Power and Energy Magazine},
  volume={21},
  number={5},
  pages={68--77},
  year={2023},
  doi={10.1109/MPE.2023.3296604}
}

@article{synchrowaveforms_2025,
  author={Hamed Mohsenian-Rad and Mladen Kezunovic and Farnoosh Rahmatian},
  title={Synchro-Waveforms in Wide-Area Monitoring, Control, and Protection: Real-World Examples and Future Opportunities},
  journal={IEEE Power and Energy Magazine},
  volume={23},
  number={1},
  pages={69--80},
  year={2025},
  doi={10.1109/MPE.2024.3471235}
}

@article{IBR_waveform_analysis_2025,
  author={Hossein Mohsenzadeh-Yazdi and Chester (Chun) Li and Hamed Mohsenian-Rad},
  title={{IBR} Responses During a Real-World System-Wide Disturbance: Synchro-Waveform Data Analysis, Pattern Classification, and Engineering Implications},
  journal={IEEE Transactions on Smart Grid},
  year={2025},
  note={Accepted for publication}
}

@article{khan2024grid,
  author={Musa Khan and Wenchuan Wu and Li Li},
  title={Grid-Forming Control for Inverter-Based Resources in Power Systems: A Review on Its Operation, System Stability, and Prospective},
  journal={IET Renewable Power Generation},
  volume={18},
  number={6},
  pages={887--907},
  year={2024},
  doi={10.1049/iet-rpg.2024.0212}
}

@article{arevalo2025systematic,
  author={Paul Arévalo and Carlos Ramos and Agostinho Rocha},
  title={A Systematic Review of Grid-Forming Control Techniques for Modern Power Systems and Microgrids},
  journal={Energies},
  volume={18},
  number={14},
  pages={3888},
  year={2025}
}

@article{hakam2026grid,
  author={Youness Hakam and Mohamed Tabaa},
  title={Grid-Forming Inverters in Photovoltaic Power Systems: A Comprehensive Review of Modeling, Control, and Stability Perspectives},
  journal={Energies},
  volume={19},
  number={5},
  pages={1244},
  year={2026}
}

@article{musca2023wide,
  author={Musca, Rossano and Sanseverino, Eleonora Riva and Zizzo, Gaetano and Giannuzzi, Giorgio and Pisani, Cosimo},
  title={Wide-synchronization control for power systems with grid-forming converters},
  journal={IEEE Transactions on Power Systems},
  volume={39},
  number={3},
  pages={4998--5007},
  year={2023},
  doi={10.1109/TPWRS.2023.3334876}
}

@article{kuruva2025robust,
  title={Robust Loop-Shaping Control for Medium-High Frequency Oscillations Mitigation in Grid-Connected Converters with Effective Hardware Implementation},
  author={Kuruva, Veeranna and Abolmasoumi, Amir H and Thomas, Vinu and Marinescu, Bogdan},
  journal={IEEE Transactions on Power Delivery},
  year={2025},
  publisher={IEEE}
}

@article{damping_low_frequency_GFM,
  title={Damping low-frequency oscillations in power systems using grid-forming converters},
  author={Rodr{\'\i}guez-Amenedo, Jos{\'e} Luis and G{\'o}mez, Santiago Arnaltes},
  journal={IEEE Access},
  volume={9},
  pages={158984--158997},
  year={2021},
  publisher={IEEE}
}

@phdthesis{mukundan2025impedance,
  title={Impedance-based Stability Analysis on IBR Integrated Power System},
  author={Mukundan, Vani},
  year={2025},
  school={Open Access Te Herenga Waka-Victoria University of Wellington}
}

@article{reviewing_GFM_paradigms,
  title={Reviewing control paradigms and emerging trends of grid-forming inverters—A comparative study},
  author={Rahman, K. and others},
  journal={Energies},
  volume={17},
  number={10},
  pages={2400},
  year={2024},
  publisher={MDPI}
}

@article{wideband_oscillation_estimation,
  title={Power system wideband oscillation estimation, localization, and mitigation},
  author={An, S. and others},
  journal={IET Generation, Transmission \& Distribution},
  volume={17},
  number={11},
  pages={2655--2666},
  year={2023},
  publisher={Wiley Online Library}
}

@misc{matlab_sps,
  author       = {{MathWorks}},
  title        = {Simscape Electrical Specialized Power Systems},
  year         = {2026},
  howpublished = {\url{https://www.mathworks.com/help/sps/specialized-power-systems.html}},
  note         = {Accessed: 2026-04-19}
}
\vspace{-0.3cm}

\end{document}